\documentstyle[12pt]{article}
\begin{document}

%\draft
\def\be{\begin{equation}}
\def\ee{\end{equation}}
\def\bea{\begin{eqnarray}}
\def\eea{\end{eqnarray}}
\def\nn{\nonumber}
\def\ep{\epsilon}
\def\c{\cite}
\def\m{\mu}
\def\ga{\gamma}
\def\lan{\langle}
\def\ran{\rangle}
\def\Ga{\Gamma}
\def\thet{\theta}
\def\la{\lambda}
\def\Lam{\Lambda}
\def\ka{\chi}
\def\rt{r^{2}}
\def\at{a^{2}}
\def\st{\sin\theta}
\def\stt{\sin^{2}\theta}
\def\ct{\cos\theta}
\def\ctt{\cos^{2}\theta}

\def\si{\sigma}
\def\al{\alpha}
\def\pa{\partial}
\def\de{\delta}
\def\De{\Delta}
\def\Dex{\Delta_{,x}}
\def\Dey{\Delta_{,y}}
\def\Dev{\Delta^{-1}}
\def\Ome{\Omega}
\def\Om2{\Omega^{2}}
\def\ov{\over}
\def\gmn{g_{\mu\nu}}
\def\gmnv{g^{\mu\nu}}

\def\rsr{{r_{s}\over r}}
\def\rrs{{r\over r_{s}}}   
\def\rs2r{{r_{s}\over 2r}}
\def\l2r2{{l^{2}\over r^{2}}}
\def\rsa{{r_{s}\over a}}
\def\rsb{{r_{s}\over b}}
\def\rsro{{r_{s}\over r_{o}}}
\def\rss{r_{s}}
\def\a2{{l^{2}\over a^{2}}}
\def\b2{{l^{2}\over b^{2}}}
\def\op{\oplus}
\def\sn{\stackrel{\circ}{n}}
\def\c{\cite}

%\twocolumn[\hsize\textwidth\columnwidth
%\hsize\csname @twocolumnfalse\endcsname

\title
{Dirac Spin Precession in Kerr Spacetime by  the parallelism 
description}
\author{C. M. Zhang\\ National Astronomical Observatory,
Chinese Academy of Sciences\\ Beijing 100012, China, 
zhangcm@bao.ac.cn } 

%\begin{center}
%\end{center}

%\date{\today}

\maketitle

\begin{abstract}

In the framework of parallelism general relativity (PGR), 
the Dirac particle spin precession in the rotational gravitational field is studied. 
 In terms of the  equivalent tetrad of Kerr  frame, we investigate  the 
torsion axial-vector  spin coupling in PGR. In the case of the 
weak field and slow rotation approximation, 
 we obtain that the torsion axial-vector has  the dipole-like  structure, 
but different from the gravitomagnetic field, 
which indicates that the choice of the Kerr tetrad will influence 
on the physics interpretation  of the  axial-vector spin coupling.

\end{abstract}

%\pacs{04.25.Nx, 04.80.Cc, 04.50.+h, 04.20.Jb}

{\bf key words}: torsion,  parallelism, Kerr spacetime, spin

%\vskip1pc

\section{Introduction}

As an extension of Einstein's general relativity, the metric-affine theory  
is one of the choices~\c{h76,h91,hmpd}, 
which  incorporates  the geometric structure of spacetime into 
curvature  and torsion. 
  In order for the fundamental exploration of spacetime, 
the tetrad theory of gravitation has been paid more attention by many 
people   ~\cite{hay79,nh80,per1,per2,perbook,pvz01,mal},  
 where the spacetime  is characterized by the
 torsion tensor and the vanishing  curvature, the relevant spacetime is the
Weitzenb\"ock spacetime~\cite{hay79}, which is a special case of the Riemann-Cartan
spacetime, such as the  metric-affine theory of
 gravitation \c{h76,h91,hmpd}. 
 The tetrad theory of gravitation  will be equivalent to general relativity
 when the convenient choice of the parameters of  the Lagrangian ~\c{hay79}.

We will use the greek alphabet ($\mu$, $\nu$, $\rho$,~$\cdots=1,2,3,4$)
to denote tensor indices, that is, indices related to spacetime. The latin alphabet
($a$, $b$, $c$,~$\cdots=1,2,3,4$) will be used to denote local Lorentz (or tangent space)
indices. Of course, being of the same kind, tensor and local Lorentz indices can be
changed into each other with the use of the tetrad $h^{a} {}_{\mu}$, which satisfy
\be
h^{a}{}_{\mu} \; h_{a}{}^{\nu} = \delta_{\mu}{}^{\nu} \quad
; \quad h^{a}{}_{\mu} \; h_{b}{}^{\mu} = \delta^{a}{}_{b} \; .
\label{orto}
\ee
A nontrivial tetrad field can be used to 
define the linear Cartan connection\c{hay79,perbook}
 
\be
\Gamma^{\sigma}{}_{\mu \nu} = h_a{}^\sigma \partial_\nu h^a{}_\mu \;,
\label{car}
\ee 
with respect to which the tetrad is parallel:  
\be {\nabla}_\nu \; h^{a}{}_{\mu}
\equiv \partial_\nu h^{a}{}_{\mu} - \Gamma^{\rho}{}_{\mu \nu}
\, h^{a}{}_{\rho} = 0 \; . 
\label{weitz}
\ee 
The Cartan connection can be decomposed according to 
\be
{\Gamma}^{\sigma}{}_{\mu \nu} = {\stackrel{\circ}{\Gamma}}{}^{\sigma}{}_{\mu
\nu} + {K}^{\sigma}{}_{\mu \nu} \; ,
\label{rel} 
\ee
where
\be
{\stackrel{\circ}{\Gamma}}{}^{\sigma}{}_{\mu \nu} = \frac{1}{2}
g^{\sigma \rho} \left[ \partial_{\mu} g_{\rho \nu} + \partial_{\nu}
g_{\rho \mu} - \partial_{\rho} g_{\mu \nu} \right]
\label{lci}
\ee
is the Levi--Civita connection of the metric

\be
g_{\mu \nu} = \eta_{a b} \; h^a{}_\mu \; h^b{}_\nu \; ,
\label{gmn}
\ee
where $\eta^{ab}$ is the metric in flat space with the line element

\be
d\tau^{2} = g_{\mu \nu} dx^{\mu} dx^{\nu} \;,
\ee 
and 
\be {K}^{\sigma}{}_{\mu \nu} = \frac{1}{2}
\left[ T_{\mu}{}^{\sigma}{}_{\nu} + T_{\nu}{}^{\sigma}{}_{\mu}
- T^{\sigma}{}_{\mu \nu} \right]
\label{conto}
\ee 
is the contorsion tensor, with 
\be
T^\sigma{}_{\mu \nu} =
\Gamma^ {\sigma}{}_{\mu \nu} - \Gamma^{\sigma}{}_{\nu \mu} \;  \label{tor} 
\ee
the torsion of the Cartan connection~\c{hay79,perbook}. The irreducible
torsion vectors, i.e., the torsion vector and 
the torsion axial-vector, can then be
constructed as~\c{hay79,perbook} 
\be
V_{\mu} =  T^{\nu}{}_{\nu \mu}\,,
\ee
\be
A_{\m} = {1\over 6}\ep_{\m\nu\rho\si}T^{\nu\rho\si}\,,
\ee
with $\ep_{\mu \nu \rho \sigma}$ being the completely antisymmetric
tensor normalized as $\ep_{0123}=\sqrt{-g}$  
and $\ep^{0123}=\frac{1}{\sqrt{-g}} $, where $g$ is the determinant of 
metric.

The spacetime dynamic effects on the spin is  incorporated into Dirac
equation  through the ``spin connection''  appearing in the Dirac equation
in  gravitation \cite{hay79}. 
 In Weitzenb\"ock spacetime, as well as the general version of torsion 
gravity, it has been shown by many
authors~\c{hay79,nh80,ham94,ham95,heh71,tra72,rum79,yas80,aud81} that the spin
precession of a Dirac particle  is intimately
related to the torsion axial-vector,
 and it is interesting to note  that the 
torsion  axial-vector represents the deviation of the axial symmetry
from the spherical symmetry \cite{nh80}.

\be
\frac{d{\bf S}}{dt} = - {3\ov2} \mbox{{\boldmath $A$}} \times {\bf S}, 
\label{precession1}
\ee
where ${\bf S}$ is the semiclassical spin vector of a Dirac particle, 
and $\mbox{{\boldmath $A$}}$ is the spacelike part 
of the torsion axial-vector.
Therefore, the corresponding extra Hamiltonian energy is of the form, 

\be
\de H = - {3\ov2} \mbox{{\boldmath $A$}}\cdot \mbox{\boldmath$S$}\,.
\label{ham2}
\ee

The purpose of the paper is to derive the torsion axial vector spin 
coupling in the  Kerr spacetime with the given  tetrad, which 
is performed in section 2. 
In the  weak field and slow rotation approximation,  
the analytical  expression  of the torsion 
axial-vector is obtained in section 3. 
 Discussions and conclusions are given in the last  section. 
 Throughout this paper we use the  unit with  $c=1$.

\section{The torsion axial-vector in Kerr spacetime}

The gravitational field of a rotating mass is described by the
axially symmetric stationary Kerr metric~\cite{mtw},
\be
d \tau^{2} = g_{00} dt^2 + g_{11} dr^2 + g_{22} d\theta^2 +
g_{33} d\phi^2 + 2 g_{03} d\phi\; dt ,
\ee
where
\be
g_{00} = 1 - { r_s r \ov \Sigma}; \;\;g_{11} = - { \Sigma \ov \Delta};\;\;
g_{22} = - \Sigma
\ee
\be
g_{33} =  - \left( r^2 + a^2 + {r_s r a^2 \ov  \Sigma} \sin^2{\theta} \right)
\sin^2{\theta}
\ee
\be
 g_{03} = g_{30} =  {r_s r a \ov \Sigma }\sin^2{\theta}
\ee
with
\be
\Delta  = r^2 - r_s r + a^2 \;\; {\rm and} \;\;
\Sigma = r^2 + a^2 \cos^2\theta \; .
\ee
In these expressions, $\rss$ is Schwarzschild radius and 
$a$ is the angular momentum of a gravitational  unit
mass source. If $a=0$, the Kerr metric becomes the Schwarzschild metric in
the standard form. In Kerr spacetime, the tetrad can be expressed by the 
dual basis of the differential one-form \c{ho} through  
choosing a coframe of the  coordinate  system, 
\bea
d\vartheta ^{\hat{0}}& =& \,  A[d\,t - a \sin^{2}\theta{} {d\phi}]\;, \\
\label{coframe0}
\quad d\vartheta ^{\hat{1}} &=&\,A^{-1}{} dr\;,\\
\quad d\vartheta ^{\hat{2}} &=&\,  \sqrt{\Sigma} d \theta\;, \\
\quad d\vartheta ^{\hat{3}} &=&\,  B[ (-a dt + (r^{2} + a^{2})d \phi]\;,
\label{coframe}
\eea
where $A=\sqrt{\Delta/\Sigma}$, $B=\sin\theta/\sqrt{\Sigma}$.
%, $\Delta = \rt + \at -\rss r$,$\Sigma= \rt + \at \cos^2\theta$.
 Therefore, the tetrad can be obtained with the subscript 
$\mu$ denoting the column index (c.f. \c{zcm2002}),  

\be
h^{a}{}_{\mu} = \pmatrix{
A   & 0 & 0 & -aA\sin^{2}\theta \cr
0 & 1/A & 0 & 0 \cr
0 & 0 & \sqrt{\Sigma}  & 0 \cr
-aB & 0 & 0 & (\rt + \at)B} ,
\label{te1}
\ee 
 with the inverse 
\be
h h_{a}{}^{\mu} = \pmatrix{
(\rt+\at)\st/A & 0  & 0 & a\st/A \cr
0 & A\Sigma\st &0  &0\cr
0&0 &B \Sigma & 0 \cr
a \sqrt{\Sigma} \stt &0 &0 &\sqrt{\Sigma} } ,
\label{te2}
\ee
where $h=\det (h^{a}{}_{\mu})= \Sigma \st =\sqrt{-g}$ 
with  $g=\det (g_{\mu \nu})$.
 We can inspect that 
Eqs.(\ref{te1}) and (\ref{te2}) satisfy the conditions in Eq.(\ref{orto}). 
 {}From Eqs.(\ref{te1}) and (\ref{te2}), we can now construct the
Cartan connection, whose nonvanishing components are:
\bea
h\Ga^{0}{}_{01} &=& (A^{'}/A) (\rt+\at)\st - B^{'} \at \sqrt{\Sigma} \stt, \\
h\Ga^{0}{}_{31} &=& -(A^{'}/A)a  (\rt+\at) \sin^{3}\theta + [B(\rt+\at)]^{'} a \sqrt{\Sigma}\stt, \\
h\Ga^{0}{}_{02} &=&  (A^{'}_{\theta}/A)(\rt+\at) \st -\at \stt \ct=a^{4}\sin^{4}\theta \ct/\Sigma, \\
h\Ga^{0}{}_{32} &=&  -[A\sin^{2}\theta]^{'}_{\theta} a\st(\rt+\at)/A + B^{'}_{\theta} 
(\rt+\at) a\sqrt{\Sigma}\sin^{2}\theta, 
\eea

\be
h\Ga^{1}{}_{12} =  -\at \stt \ct, \quad  h\Ga^{2}{}_{21} = r \st\;,
\ee

\bea
h\Ga^{3}{}_{01} &=& (A^{'}/A)a\st  - B^{'} a \sqrt{\Sigma},  \\
h\Ga^{3}{}_{31} &=& -(A^{'}/A)\at \sin^{3}\theta  + [B(\rt+\at) ]^{'} \sqrt{\Sigma}, \\
h\Ga^{3}{}_{02} &=& (A^{'}_{\theta}/A)a\st  - B^{'}_{\theta} \sqrt{\Sigma} a = (\at/\Sigma-1)a\ct,  \\
h\Ga^{3}{}_{32} &=&  -(A\sin^{2}\theta)^{'}_{\theta} \at \st/A  + B^{'}_{\theta} (\at+\rt) \sqrt{\Sigma}\;,
\eea 
where 
\bea
A^{'}&=&{\pa A \ov \pa r} = (r-\rss/2)/(A\Sigma)-Ar/\Sigma,  
A^{'}_{\theta}= {\pa A \ov \pa \theta} = {\at A\ov \Sigma}\st\ct\;, \\ 
B^{'}&=&{\pa B\ov \pa r}= -Br/\Sigma\;, 
B^{'}_{\theta}={\pa B\ov \pa \theta} = \ct/\sqrt{\Sigma} + {\at B\ov \Sigma} \st\ct\;.
\eea

The nonzero  torsion axial--vector components are
\bea
\label{a1}
A^{(1)}\times(6h) &=& -2 (g_{00}T^{0}{}_{23} + g_{03}T^{3}{}_{23} 
+ g_{30}T^{0}{}_{02} + g_{33}T^{3}{}_{02} ) \\\nn
 &=& -2 (g_{00}\Ga^{0}{}_{32} + g_{03}\Ga^{3}{}_{32} 
- g_{30}\Ga^{0}{}_{02} - g_{33}\Ga^{3}{}_{02} )\; , 
\eea

\bea
\label{a2}
A^{(2)}\times(6h) &=& 2[g_{00}T^{0}{}_{13} + g_{03}T^{3}{}_{13} +
g_{30}T^{0}{}_{01}+g_{33}T^{3}{}_{01}] \\ \nn
&=& 2[g_{00}\Ga^{0}{}_{31} + g_{03}\Ga^{3}{}_{31} -
g_{30}\Ga^{0}{}_{01}-g_{33}\Ga^{3}{}_{01}] \;.
\eea

\section{Slow Rotation and Weak Field Approximations}
                                                                                                                     
In the case of slow rotation and weak field, 
we keep the terms up to first order in the
angular momentum $a$ and in $\rss/r$. 
The related quantities are simplified as follows:
\be
\Delta  = r^2 - r_s r; \quad \Sigma = r^2
\label{delta}
\ee
\be
g_{00}=(-g_{11}) ^{-1} = 1 - { r_s  \ov r}; \quad g_{22}= - r^2
\label{g00}
\ee
\be
g_{33}= - r^2 \sin^2{\theta}; \quad g_{03} = {r_s a \ov r }\sin^2{\theta} 
\label{g33}
\ee

\be 
h=r^2 \sin\theta; \quad A=\sqrt{1-\rss/r}; \quad B=\st/r\;. 
\label{happ}
\ee
In this approximation, all terms  reduce to
the values of the Schwarzschild solution except  $g_{03}$. 
On the other hand, in the weak field limit,
characterized by keeping terms up to first order in $\rss/r$, the nonzero components
of the axial--vector torsion become

\bea
&&h\Ga^{0}{}_{32} = -a\rt \stt \ct\;,  \quad  h\Ga^{3}{}_{32} = \rt \ct, \nn\\ 
&&h\Ga^{0}{}_{02} = 0\;, \quad h\Ga^{3}{}_{02}  = - a\ct\;,
\eea
and 
\bea
&&h\Ga^{0}{}_{31} = a r \sin^{3}\theta (1-\rss/2r)\;,  \quad  h\Ga^{3}{}_{31} = r \st, \nn\\ 
&&h\Ga^{0}{}_{01} = (\rss/2) \st\;, \quad h\Ga^{3}{}_{01}  =  (a\st/r)(1+\rss/2r) \;.
\eea
Substituting Eqs.(\ref{delta}), (\ref{g00}), (\ref{g33}) and  (\ref{happ}) into 
(\ref{a1})  and (\ref{a2}), we obtain
\be
A^{(1)} = {2 \ov 3} (1-\rss/r) { a\ov  r^2} \, \cos{\theta}\;,
\ee
and
\be
A^{(2)} = {2\ov 3} { a\ov  r^{3}} \sin{\theta}\;.
\ee
In spacelike vector form, the axial--vector becomes,
\be
-\mbox{{\boldmath $A$}} = A^{(1)} \sqrt{-g_{11}} \; {\bf e}_{r} +
A^{(2)} \sqrt{-g_{22}} \; {\bf e}_{\theta} ,
\ee
where  ${\bf e}_{r} = \sqrt{-g_{11}}\; dr$  and 
 ${\bf e}_{\theta}= \sqrt{-g_{22}}\; d\theta $ are unit 
vectors in (r,$\;\theta$) directions, and  then we have,
\be
-\mbox{{\boldmath $A$}} = {2a\ov 3 r^{2}} [\sqrt{1-\rss/r}\; \cos{\theta} \; {\bf e}_{r}
+ \sin{\theta} \; {\bf e}_{\theta} ] .
\label{av1}
\ee
 It has been shown by many
authors~\c{hay79,ham95,heh71,tra72} that the spin
precession of a Dirac particle in torsion gravity is intimately related to
the axial--vector
\be
\frac{d{\bf s}}{dt} = {\bf b} \times {\bf s}\;,
\ee
where ${\bf s}$ is the spin vector, and ${\bf b} = - 3 \mbox{{\boldmath $A$}}/2$.
Therefore,
\be
{\bf b} = {J\ov M r^{2}} [\sqrt{1-\rss/r}\; \cos{\theta} \; {\bf e}_{r}
+ \sin{\theta} \; {\bf e}_{\theta} ] .
\label{av2}
\ee
with $J=Ma$ the angular momentum.

\section{Discussions and conclusions}

The torsion axial-vector  Dirac spin coupling by the special choice of the 
Kerr tetrad in the framework of the 
 torsion  spacetime without curvature has been derived. 
We employ the given Kerr tetrad  to derive the torsion 
axial-vector, as one of the 
three  irreducible quantities in Weitzenb\"ock spacetime, which will 
couple with  the Dirac spin. Unlike the previous work where another Kerr 
tetrad is used ~\c{pvz01}, we have  not 
obtained the gravitomagnetic spin coupling here, 
which indicates that the choice of the 
tetrad incurs the preferred 
reference frame where the physics measurement is performed.  
  It is worth noting the 
implications of some special cases from the analysis of the 
${\bf b}$ field in Eq.(\ref{av2}), and we find that 
 ${\bf b}$ field is still a dipole-like field although 
it is not a standard 
dipole gravitomagnetic field as obtained before ~\c{pvz01}, which 
is on  account of the axisymmetric property of Kerr  spacetime.
 If we set the gravitational constant G=0 or $\rss=0$, 
say the null gravitational field, then we find that ${\bf b}$ 
field is not vanished. This fact shows that G=0 will arise 
the spacetime  curvature to be zero, but the torsion would 
not be  cancelled automatically. The similar phenomenon has 
also been found when we deal with  the rotation spin coupling 
in the  rotational system, where the nonzero Cartan connections 
still survive  in the Minkowski spacetime ~\c{zcm2003}.   
 Of course, if there is no rotation, say a=0, then the 
${\bf b}$ field disappears because the axial-vector represents 
 the measurement of the axial symmetry deviated from the spherical 
symmetry~\c{nh80}.  

%\section*{Acknowledgments}This work has been supported by NSF of China.

\end{document}